\documentclass{phb-proc4-auth}

\usepackage{graphicx}
\usepackage{amssymb}

\newcommand{\mib}[1]{\mbox{\boldmath $#1$}} 

\begin{document}
\begin{frontmatter}


\journal{SCES '04}


\title{Orbital ordering and one-dimensional magnetic correlation 
in vanadium spinel oxides $A$V$_2$O$_4$ ($A$ = Zn, Mg, or Cd)}

%
%
%
%
%
%

\author[RIKEN]{Yukitoshi Motome\thanksref{ABC}\corauthref{1}}
\author[YUKAWA]{Hirokazu Tsunetsugu\thanksref{ABC}}

%
 
\address[RIKEN]{RIKEN (The Institute of Physical and Chemical Research),
Saitama 351-0198, Japan}
\address[YUKAWA]{Yukawa Institute for Theoretical Physics, Kyoto University, 
Kyoto 606-8502, Japan}

%
%
%
%

\thanks[ABC]{This work is supported by a Grant-in-Aid and NAREGI 
from the Ministry of Education, Science, Sports, and Culture. 
A part of the work was accomplished during Y. M. was staying 
at the Yukawa Institute of Theoretical Physics, 
with the support from The 21st Century for Center of Excellence program, 
`Center for Diversity and Universality in Physics'.}

%
%
%
%

\corauth[1]{Corresponding Author: RIKEN, 2-1 Hirosawa, Wako, 
Saitama 351-0198, Japan. Phone: +81-48-467-1379, 
Fax: +81-48-467-1339, Email: motome@riken.jp}


\begin{abstract}

We present our theoretical results on the mechanism of two transitions 
in vanadium spinel oxides $A$V$_2$O$_4$ ($A$=Zn, Mg, or Cd)
in which magnetic V cations constitute a geometrically-frustrated 
pyrochlore structure. 
We have derived an effective spin-orbital-lattice coupled model 
in the strong correlation limit of the multiorbital Hubbard model, and  
applied Monte Carlo simulation to the model. 
The results 
reveal that the higher-temperature transition is 
a layered antiferro-type orbital ordering 
accompanied by tetragonal Jahn-Teller distortion, and 
the lower-temperature transition is an antiferromagnetic spin ordering. 
The orbital order lifts the magnetic frustration partially, and 
induces spatial anisotropy in magnetic exchange interactions. 
In the intermediate phase, the system can be considered to consist of 
weakly-coupled antiferromagnetic chains 
lying in the perpendicular planes to the tetragonal distortion. 

\end{abstract}

%
%



\end{frontmatter}

%
%
%
%
%

Vanadium spinel oxides $A$V$_2$O$_4$ with nonmagnetic cations 
$A$=Zn, Mg, or Cd are one of the most typical geometrically-frustrated systems; 
the magnetic V cations constitute the pyrochlore lattice, 
which is a three-dimensional network of corner-sharing tetrahedra. 
The geometrical frustration strongly suppresses 
development of magnetic correlations, and 
the antiferromagnetic (AF) ordering sets in at a much lower temperature $\sim 40$K 
than the Curie-Weiss temperature $\sim 1000$K
\cite{Ueda1997}.
Besides the AF transition, at a slightly higher temperature $\sim 50$K, 
the system shows another phase transition 
with the structural change from high-temperature cubic 
to low-temperature tetragonal symmetry
\cite{Ueda1997}.
The origin of the two successive transitions is under debate. 
In particular, theories based on spin degree of freedom only or 
the spin-lattice coupling appear to be insufficient 
to describe the two transitions in a comprehensive manner
\cite{Tsunetsugu2003}.

In the present study, 
we focus on the $t_{2g}$ orbital degree of freedom in these vanadium spinels 
to understand the microscopic mechanism of the two successive transitions. 
Each V$^{3+}$ cation has two $3d$ electrons 
in threefold $t_{2g}$ orbitals, and 
therefore the orbital degree of freedom is active. 
Starting from the standard multiorbital Hubbard model, 
we consider the perturbation in the strong correlation limit to describe 
the low energy physics of these insulating materials, and 
derive the effective spin-orbital-lattice coupled model in the form 
\cite{Tsunetsugu2003}
\begin{eqnarray}
&& H = H_{\rm SO} + H_{\rm JT}, 
\label{eq:H}
\\
&& H_{\rm SO} = -J \sum_{\langle i,j \rangle} h_{ij} 
- J_3 \sum_{\langle\!\langle i,j \rangle\!\rangle} h_{ij}, 
\label{eq:H_SO}
\\
&& h_{ij} = (A + B \mib{S}_i \cdot \mib{S}_j) 
[ n_{i \alpha(ij)} \bar{n}_{j \alpha(ij)} 
+ \bar{n}_{i \alpha(ij)} n_{j \alpha(ij)} ]
\nonumber \\
&& \quad \quad + C (1 - \mib{S}_i \cdot \mib{S}_j)
n_{i \alpha(ij)} n_{j \alpha(ij)}, 
\\
&& H_{\rm JT} = \gamma \sum_i Q_{i} \epsilon_i
+ \sum_i Q_i^2 / 2 
- \lambda \sum_{\langle i,j \rangle} Q_i Q_j, 
\end{eqnarray}
where $\mib{S}_i$ is the $S=1$ spin operator and  
$n_{i \alpha}$ 
is the density operator for site $i$ and 
orbital $\alpha = 1$ ($d_{yz}$), $2$ ($d_{zx}$), $3$ ($d_{xy}$). 
Here, $\bar{n}_{i\alpha} = 1 - n_{i\alpha}$ and 
$\epsilon_i = n_{i1} + n_{i2} - 2 n_{i3}$; and 
we impose a local constraint $\sum_{\alpha=1}^3 n_{i \alpha} = 2$ at each site. 
The summations with $\langle i,j \rangle$ and 
$\langle\!\langle i,j \rangle\!\rangle$ 
are taken over the nearest-neighbor (NN) sites and third-neighbor sites, respectively. 
Here, 
we take into account only the dominant $\sigma$-bond hopping integrals 
in the original multiorbital Hubbard model, which results in 
the orbital diagonal interaction in $H_{\rm SO}$; 
$\alpha(ij)$ is the orbital which gives rise to the $\sigma$ bond 
between sites $i$ and $j$. 
$H_{\rm JT}$ describes the orbital-lattice coupling part, 
where $\gamma$ is the electron-phonon coupling constant 
of the tetragonal Jahn-Teller (JT) mode, 
$Q_i$ denotes the amplitude of local lattice distortion at site $i$, and 
$\lambda$ describes the interaction between NN JT distortions, 
which mimics the cooperative aspect of the JT distortion. 
The parameters in $H_{\rm SO}$ are given by the coupling constants 
in the starting multiorbital Hubbard Hamiltonian, and 
we use the reasonable estimates as 
$J_3/J = 0.02$ with $J \simeq 200$K, 
$A = 1.21$, $B = 0.105$, and $C = 0.931$.
For the JT parameters, we take $\gamma^2/J = 0.04$ and $\lambda/J = 0.15$, 
which are typical values to reproduce the tetragonal distortion 
observed in experiments. 

We have performed Monte Carlo (MC) simulation to investigate 
thermodynamic properties of the model (\ref{eq:H}), 
which is a classical simulation to avoid the negative sign problem 
due to the geometrical frustration. 
Since quantum nature exists only in the spin operators in $H_{\rm SO}$, 
we approximate them by classical vectors, and 
apply a standard metropolis MC algorithm. 
Details of MC calculations will be reported elsewhere
\cite{MotomePREPRINT}. 

\begin{figure}
\includegraphics[width=7cm]{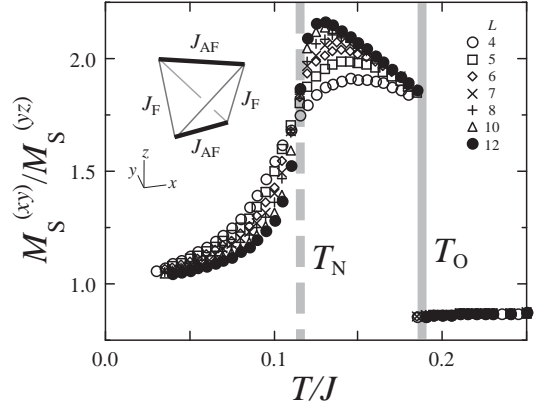}
\caption{
Monte Carlo results of the anisotropy between the staggered moments 
in the $xy$ chains and in the $yz$ chains. 
The orbital ordering temperature $T_{\rm O}$ 
and the antiferromagnetic spin ordering temperature $T_{\rm N}$ are shown. 
The system size is $L^2 \times 16$ sites.
Inset: Anisotropic spin exchange interactions under the orbital ordering. 
} 
\label{fig} 
\end{figure}

MC results show that the model (\ref{eq:H}) exhibits two transitions:
One is first order transition at $T_{\rm O} \simeq 0.19J$  
corresponding to the orbital ordering with the tetragonal JT distortion, and  
the other is at $T_{\rm N} \simeq 0.115J$ 
with continuous growth of the AF spin ordering. 
The orbital order below $T_{\rm O}$ is a layered antiferro type; 
($d_{xy}$, $d_{yz}$) and ($d_{xy}$, $d_{zx}$)-occupied planes 
stack alternatively in the $c$ direction 
($c$ is the axis of the tetragonal distortion). 
This orbital ordering plays a crucial role 
to reduce the magnetic frustration in the following way: 
Let us consider an effective spin Hamiltonian 
by replacing the orbital part in $H_{\rm SO}$ by its mean-field. 
The model has highly anisotropic spin exchanges; 
for instance, between NN sites, 
the strong AF interaction $J_{\rm AF} \simeq 0.931J$ in the $xy$ direction 
whereas the weak ferromagnetic interaction $J_{\rm F} \simeq -0.105J$ 
in the $yz$ and $zx$ directions. 
Moreover, the weak ferromagnetic interactions $J_{\rm F}$ are completely frustrated 
between the AF $xy$ chains 
due to the geometry of the pyrochlore structure. 
These anisotropic exchange interactions are shown in the inset of 
Fig.~\ref{fig} in a tetrahedron unit. 
As a consequence, the magnetic correlation becomes highly anisotropic, and 
the system looks like a weakly-coupled AF chains. 
Figure~\ref{fig} shows 
the ratio between the staggered moments along the $xy$ chains and 
along the $yz$($zx$) chains. 
Note that the latter moment corresponds to the ordering structure 
with periodicity four along the $yz$($zx$) chains. 
Above $T_{\rm N}$, both moments are zero in the limit of $L \to \infty$, 
and therefore the ratio gives information on 
the anisotropy of the magnetic correlation length. 
The result shows that the anisotropy is suddenly enhanced below $T_{\rm O}$ and 
takes a large value for $T_{\rm N} < T < T_{\rm O}$. 

Below $T_{\rm N}$, the AF $xy$ chains are aligned 
by the third-neighbor interaction $J_3$ 
to form the three-dimensional AF order, 
whose pattern is consistent with the neutron scattering results
\cite{Niziol1973}. 
There, the ratio in Fig.~\ref{fig} approaches 1 
in the limit of $L \to \infty$ 
since both moments are finite and the same. 

The one-dimensional anisotropy in the intermediate phase $T_{\rm N} < T < T_{\rm O}$ 
can be observed, for example, 
by neutron scattering measurement. 
Such experiment is desired to confirm our scenario.

%
%
%
%

%
%
%
%


\end{document}